\documentclass[twocolumn,aps,prc,floatfix]{revtex4}
\usepackage{graphicx}
\usepackage{dcolumn}
\usepackage{bm}
\usepackage[version=4]{mhchem}
\usepackage{color}
\def\es0{$E_{\rm sym}(\rho_0)$}

\def\us0{$U_{\rm sym}(\rho_0,k_F)$~}

\def\l0{$L(\rho_0)$~}

\begin{document}

\title{Illuminating Dark Matter Admixed in Neutron Stars with Simultaneous Mass–Radius Constraints}

\author{Nai-Bo Zhang$^{1}$\footnote{naibozhang@seu.edu.cn}, Bao-An Li$^2$\footnote{Bao-An.Li@Tamuc.edu}, Jia-Yu Zhang, Wei-Na Shen, and Hui Zhang}

\affiliation{$^1$School of Physics, Southeast University, Nanjing 211189, China}
\affiliation{$^2$Department of Physics and Astronomy, Texas A$\&$M University-Commerce, Commerce, TX 75429-3011, USA}
\date{\today}

\setcounter{MaxMatrixCols}{10}

\begin{abstract}
We investigate how simultaneous mass and radius measurements of massive neutron stars can help constrain the properties of dark matter possibly admixed in them. Within a fermionic dark matter model that interacts only through gravitation, along with a well-constrained nuclear matter equation of state, we show that the simultaneous mass and radius measurement of PSRJ0740+6620 reduces the uncertainty of dark matter central energy density by more than 50\% compared to the results obtained from using the two observables independently, while other dark matter parameters remain unconstrained. Additionally, we find that the dark matter fraction $f_D$ should be smaller than 2\% when constrained by the observed neutron star maximum mass alone, and it could be even smaller than 0.3\% with the simultaneous measurement of mass and radius, supporting the conclusion that only a small amount of dark matter exists in dark matter admixed neutron stars (DANSs).
\end{abstract}
\maketitle


\section{Introduction}

The existence of dark matter has been widely supported through observations across multiple cosmic scales, such as galactic rotation curves~\cite{Vera70}, gravitational lensing~\cite{Douglas06}, cosmic microwave background anisotropies~\cite{Planck20}, and large-scale structure formation~\cite{White78}. The considerable fraction of dark matter, approximately 26.8\% of the total energy content of the universe, undoubtedly plays a crucial role in shaping the large-scale structure of the universe~\cite{Jarosik11}. However, despite its prevalence, the properties of dark matter remain elusive. Various models and candidates have been proposed, spanning a vast mass range from $10^{-25}$ eV to $10^{25}$ eV~\cite{Baryakhtar22}. These include weakly interacting massive particles (WIMPs)~\cite{Kouvaris11,Steigman12,Bertone18}, axions~\cite{Duffy09,Olive14,Tanabashi18}, dark photons~\cite{An15}, sterile neutrinos~\cite{Bertone18b}, and mirror matter~\cite{Blinnikov82,Blinnikov83,Khlopov91}, etc.

Direct detection experiments have attempted to constrain the properties of dark matter, with significant contributions from detectors like XENON1T~\cite{Aprile18,Aprile19,Aprile19b,Aprile19c,Aprile20,Aprile21}, DAMA/LIBRA~\cite{Bernabei10}, LUX-ZEPLIN~\cite{Akerib14}, CRESST~\cite{Angloher16,Abdelhameed19}, CDMSLite~\cite{Agnese14,Agnese18}, CYGNUS~\cite{Vahsen20}, and PANDAX-II~\cite{Wang20}. However, no conclusive results have been obtained at present.

Alternatively, dark matter properties can be constrained indirectly. Neutron stars, the densest visible objects in the universe, provide an ideal environment to study the fundamental physics of supradense matter, including the effects of dark matter. As dark matter particles pass through a neutron star, their interactions with the dense core can lead to the capture of dark matter, forming a dark matter admixed neutron star (DANS). The presence of dark matter may modify various neutron star properties and influence the gravitational wave signatures produced during binary neutron star mergers. These effects can potentially be identified through neutron star observations (see a recent review in ref.~\cite{Grippa25}). The amount of dark matter in a neutron star depends on the star's evolutionary history and its location throughout its life. Depending on the quantity of dark matter, it may accumulate in the core of the neutron star, forming a dark matter core, or it may distribute itself within and outside the star’s baryonic radius, forming a dark matter halo.

With the accumulation of more neutron star observation data, constraints on dark matter properties have become increasingly possible. For instance, the mass of the massive neutron star, PSR J0740+6620, has been constrained to $2.08 \pm 0.07$ M$_\odot$~\cite{Cromartie19,Fonseca21}, with its radius constrained to $R=13.7^{+2.6}_{-1.5}$ km~\cite{Miller21} and $R=12.39^{+1.30}_{-0.98}$ km~\cite{Riley21} by the Neutron Star Interior Composition Explorer (NICER) Collaboration. Additionally, the tidal deformability of a canonical neutron star has been extracted from the binary neutron star merger event GW170817, yielding $70 < \Lambda_{1.4} < 580$ at the 90\% confidence level~\cite{Abbott18}. Combining results from NICER and GW170817, ref.~\cite{Guha24} constrained the dark matter mass $m_D$ to the range $0.1\sim30$ MeV, while ref.~\cite{Ivanytskyi20} proposed an upper limit of 60 GeV.

Significant attention has been devoted to studying how dark matter can affect the properties of DANSs. The existence of dark matter can substantially alter the spacetime around a neutron star, impacting observable quantities such as gravitational wave signals after mergers, mass--radius profiles, tidal deformabilities, and thermal evolution~\cite{Liang12,Xiang14,Panotopoulos17,Nelson19,Ivanytskyi20,Karkevandi22,Miao22,Konstantinou24,Shakeri24,Shawqi24,Ellis18,Karkevandi24,Bastero-Gil24,Scordino25,Giangrandi25,Das21,Das21b,Kain21,Das22,Barbat24,Flores24,Giangrandi24,Kumar24,Shirke24,Rutherford25,Thakur24,Kumar25}. {The transition between a dark matter core and halo depends on the dark matter particle mass $m_D$ and mass fraction $f_D$~\cite{Grippa24}. The dark matter halos can be categorized into two types, namely a compact halo with the same order of nuclear matter radius or a diffuse halo with a radius of up to hundreds of kilometers~\cite{Shawqi24}. Among these, only compact halos with large $f_D$ can significantly alter the radius of nuclear matter, as most of dark matter mass resides within the radius of nuclear matter, exerting a substantial inward gravitational pull on the nuclear matter. Other halo types have limited effects on the radius of nuclear matter.}

{In ref.~\cite{Karkevandi22}, the authors found that a dark matter halo forms when $m_D$ is in the range of 105$\sim$200 MeV, while a dark matter core forms if $m_D$ exceeds 200 MeV. Similarly, other studies~\cite{Ivanytskyi20,Giangrandi24} confirmed that a dark matter halo is present for dark matter masses in the range of a few hundred MeV. Moreover, the presence of a dark matter core could result in observable features in gravitational wave signals, such as an additional peak or an increased likelihood of prompt collapse within the first 2 ms after a merger~\cite{Ellis18,Giangrandi25}. More importantly, dark matter halos can modify the pulse profiles of neutron stars, which in turn affects the interpretation of NICER observations. Given that our study employs simultaneous mass--radius constraints from NICER to constrain dark matter properties, we focus on dark matter cores to ensure self-consistency in our analysis.}

In this study, we focus on fermionic dark matter, where gravity is the only interaction between the dark matter and nuclear matter~\cite{Shapiro83,Glendenning97,Xiang14,Ellis18,Collier22,Das22b,Routaray23,Grippa24}. The equation of state (EOS) for nuclear matter is taken from ref.~\cite{Zhang25}. It satisfies all the current observational constraints on the neutron star maximum mass as well as the radii and masses of several neutron stars by NICER and GW170817 results. Unlike previous studies, we systematically vary dark matter EOS parameters and examine how the simultaneous measurements of the neutron star mass and radius of PSRJ0740+6620 can help constrain the properties of fermionic dark matter. The rest of the paper is organized as follows: in {Section}~\ref{sec2}, we introduce the nuclear matter and dark matter EOSs; in {Section}~\ref{sec3}, we discuss in detail how the simultaneous measurement of neutron star mass and radius can constrain dark matter properties; and in {Section}~\ref{sec4}, we summarize our findings.

\section{Theoretical Framework}\label{sec2}

The EOS for both high-density nuclear matter and dark matter remains highly uncertain. For instance, refs.~\cite{Grippa25,Burgio21} summarize the possible EOSs for both nuclear matter and dark matter. {All the dark matter candidates could in principle be employed to study DANSs, and none can be ruled out with existing observational evidence. In the present study, we assume the dark matter to be fermionic dark matter.} We use the same meta-model EOS for nuclear matter as in ref.~\cite{Zhang25}. Once the EOSs are determined, the properties of DANSs can be obtained by solving the traditional Tolman--Oppenheimer--Volkov (TOV) equations or the two fluid TOV equations. The effects of dark matter on the properties of DANSs can be analyzed either through Bayesian methods~\cite{Rutherford25,Liu25}, which can handle statistically many parameters and provide insights into their interrelations, or by selecting representative parameter sets to model DANSs directly~\cite{Xiang14,Routaray23,Grippa24}. The Bayesian analysis approach allows for a probabilistic view of the parameter space, while the use of representative parameter sets offers a direct connection between the observational data and specific model parameters.

\subsection{The EOS of Nuclear Matter}\label{2.1}

In ref.~\cite{Zhang18}, we constructed a parameterized EOS to describe the $npe\mu$ matter at $\beta$-equilibrium in neutron stars. This was achieved by separately parameterizing the EOS for symmetric nuclear matter (SNM) $E_0(\rho)$ and nuclear symmetry energy $E_{\rm sym}(\rho)$ as follows: 
\begin{equation}\label{E0-taylor}
E_0(\rho) = E_0(\rho_0) + \frac{K_0}{2} \left(\frac{\rho - \rho_0}{3 \rho_0}\right)^2 + \frac{J_0}{6} \left(\frac{\rho - \rho_0}{3 \rho_0}\right)^3,
\end{equation}
\begin{eqnarray}\label{Esym-taylor}
E_{\rm sym}(\rho) &=& E_{\rm sym}(\rho_0) + L \left(\frac{\rho - \rho_0}{3 \rho_0}\right) \\ \nonumber
& +& \frac{K_{\rm sym}}{2} \left(\frac{\rho - \rho_0}{3 \rho_0}\right)^2+\frac{J_{\rm sym}}{6} \left(\frac{\rho - \rho_0}{3 \rho_0}\right)^3,
\end{eqnarray}
where $E_0(\rho_0)$, $K_0$, and $J_0$ are the binding energy, incompressibility, and skewness of SNM at the saturation density $\rho_0$, while $E_{\rm sym}(\rho_0)$, $L$, $K_{\rm sym}$, and $J_{\rm sym}$ are the {magnitude}, slope, curvature, and skewness of symmetry energy at $\rho_0$. 

For more details on this model, the reader can refer to our previous publications~\cite{Zhang19a, Zhang19b, Zhang20, Zhang22, Zhang25, Xie24}. In ref.~\cite{Zhang25}, we used an EOS for nuclear matter in which all parameters were constrained based on current terrestrial nuclear experiments and neutron star observations. Specifically, the most probable values of these parameters are determined to be $E_0(\rho_0)=-16$ MeV, $K_0=240$ MeV, $J_0=-190$ Mev, $E_{\rm sym}(\rho_0)=31.7$ MeV, $L=58.7$ MeV, $K_{\rm sym}=-100$ MeV, and $J_{\rm sym}=800$ MeV. This EOS, labeled as PEOS in the following discussions, features a maximum mass of $M_{\rm max} = 2.13$~M$_\odot$ and radii for stars with masses of $M = 2.08$ M$_\odot$ and $M = 1.4$ M$_\odot$ given by $R_{2.08} = 12.25$ km and $R_{1.4} = 12.73$ km, respectively. In this study, we utilize the PEOS to reduce the complexity arising from the uncertainties in the nuclear matter EOS.

\subsection{The EOS of Dark Matter}

As mentioned earlier, numerous models and candidates have been proposed to explain the nature of dark matter. In this study, we treat the dark matter particles as fermions that primarily interact through gravity. Both vector and scalar fields are considered, with a neutral scalar meson coupling to the dark matter particles as $g_s \bar{\psi}_{\mathrm{D}}\psi_{\mathrm{D}} \phi$ and a neutral vector meson coupling to the conserved dark matter current as $g_v\overline{\psi}_{\mathrm{D}}\gamma_{\mu}\psi_{\mathrm{D}}V^{\mu}$, where $g_i$ represents the coupling constants between dark matter and the mediator. Thus, similar to the potential for baryons, one can define an effective Yukawa potential for dark matter with the exchange of bosons~\cite{Fetter71}:
\begin{equation}\label{potential}
     V_{\mathrm{eff}}(r)=\frac{g_{v}^{2}}{4 \pi} \frac{e^{-m_{v} r}}{r}-\frac{g_{s}^{2}}{4 \pi} \frac{e^{-m_{s} r}}{r}.
\end{equation}
The above potential is governed by the coupling constants $g_i$ and the masses of the mediators $m_i$. With appropriate coupling constants and masses, this potential can be attractive at large distances and repulsive at short distances. 

The full Lagrangian value, including both attractive and repulsive interactions, is given by ref.~\cite{Xiang14} as
\begin{eqnarray}
\mathcal{L}_{D}&=&\bar{\psi}_{D}\left[\gamma_{\mu}\left(i \partial^{\mu}-g_{\mu} V^{\mu}\right)-\left(m_{D}-g_{s} \phi\right)\right] \psi_{D} \\ \nonumber 
& +&\frac{1}{2}\left(\partial_{\mu} \phi \partial^{\mu} \phi-m_{s}^{2} \phi^{2}\right)-\frac{1}{4} D_{\mu \nu} D^{\mu \nu}+\frac{1}{2} m_{v}^{2} V_{\mu} V^{\mu},
\end{eqnarray}
where $D_{\mu \nu}=\partial_{\mu} V_{\nu}-\partial_{\nu} V_{\mu}$ is the strength tensor for the vector meson.

Within the relativistic mean field approximation, the vector and scalar field operators are replaced by their ground state expectation values: $\langle \phi(x) \rangle = \phi_0$ and $\langle V_{\mu}(x) \rangle = V_0$. The energy density and pressure for dark matter can then be calculated as
\begin{eqnarray}\label{energy}
\varepsilon_{D}&=& \frac{2}{(2 \pi)^{3}} \int^{k_{F_{D}}} d^{3} k \sqrt{k^{2}+\left(M_{D}^{*}\right)^{2}}+\frac{g_{v}^{2}}{2 m_{v}^{2}} \rho_{D}^{2}\\  \nonumber 
&+&\frac{m_{s}^{2}}{2 g_{s}^{2}}\left(m_{D}-m_{D}^{*}\right)^{2}, 
\end{eqnarray}
\begin{eqnarray}\label{pressure}
p_{D}&=& \frac{1}{3} \frac{2}{(2 \pi)^{3}} \int^{k_{F_{D}}} d^{3} k \frac{k^{2}}{\sqrt{k^{2}+\left(M_{D}^{*}\right)^{2}}}+\frac{g_{v}^{2}}{2 m_{v}^{2}} \rho_{D}^{2}\\  \nonumber 
&-&\frac{m_{s}^{2}}{2 g_{s}^{2}}\left(m_{D}-m_{D}^{*}\right)^{2},
\end{eqnarray}
where $m_D^*=m_D-g_s\phi_0$ is the effective mass of dark matter and $\rho_D$ is the number density of dark matter.

In this dark matter model, it is found that the dark matter EOS depends only on the ratio of $C_{DV} = g_v / m_v$ and $C_{DS} = g_s / m_s$. There are no reliable constraints on the model parameters, and it is only required that $C_{DS} < C_{DV}$ to ensure that the potential is attractive at large distances and repulsive at short distances~\cite{Xiang14,Das22b}. Consequently, the parameters that affect the EOS are $m_D$, $C_{DV}$, and $C_{DS}$. The typical values of $C_{DS}$ and $C_{DV}$ range from 0 to 20 GeV$^{-1}$. For example, ref.~\cite{Das22b} suggests that $C_{DS} = 3.90^{+0.82}_{-0.70}$ GeV$^{-1}$ and \mbox{$C_{DV} = 11.88^{+0.53}_{-0.46}$ GeV$^{-1}$}. Additionally, several representative combinations of $C_{DS}$ and $C_{DV}$ are often chosen to study the effects of dark matter on the properties of DANSs. For instance, combinations like $C_{DS} = 0$ GeV$^{-1}$ and $C_{DV} = 10$ GeV$^{-1}$ and $C_{DS} = 4$ GeV$^{-1}$ and $C_{DV} = 10$ GeV$^{-1}$ have been used in refs.~\cite{Xiang14, Routaray23}. The mass of dark matter $m_D$ is crucial for determining the formation of the dark matter core or halo. Ref.~\cite{Karkevandi22} found that a dark matter halo forms when $105 < m_D < 200$ MeV, and a dark matter core forms if $m_D > 200$ MeV. The effects of varying $m_D$ are explored in ref.~\cite{Routaray23}. In this study, since reliable constraints on $m_D$ are lacking and given our focus on the study of the dark matter core, we select $m_D$ values in the range from 500 MeV to 2000 MeV. {Dark matter with $m_D<500$ MeV would lead to a dark matter halo for DANSs.}

\subsection{Two-Fluid TOV Equations}

Once the EOSs for nuclear matter and dark matter are prepared, the properties of DANSs are determined by solving the two-fluid TOV equations, given by~\cite{Ciarcellut11}
\begin{equation}
 \frac{{\rm d} P_{N}(r)}{{\rm d} r}=-\left[P_{N}(r)+\varepsilon_{N}(r)\right] \frac{4 \pi r^{3}\left[P_{N}(r)+P_{D}(r)\right]+m(r)}{r(r-2 m(r))} ,
\end{equation}
\begin{equation}
 \frac{{\rm d} P_{D}(r)}{{\rm d} r}=-\left[P_{D}(r)+\varepsilon_{D}(r)\right] \frac{4 \pi r^{3}\left[P_{N}(r)+P_{D}(r)\right]+m(r)}{r(r-2 m(r))} ,
\end{equation}
\begin{equation}
 \frac{{\rm d} m(r)}{{\rm d} r}=4 \pi\left(\varepsilon_{1}(r)+\varepsilon_{2}(r)\right) r^{2},
\end{equation}
where $r$ is the radial coordinate from the center of the star and $P_i$ and $\varepsilon_i$ represent the pressure and energy density of nuclear matter or dark matter, respectively. The total pressure and energy density are given by {$P(r) = P_D(r) + P_N(r)$} and $\varepsilon(r) = \varepsilon_D(r) + \varepsilon_N(r)$. 

To solve the above equations, initial conditions {and boundary conditions must be specified. In the center of the star ($r=0$), we set $M_N(0)=M_D(0)=0$. For the central densities of nuclear matter and dark matter, there are no reliable constraints on the central density of dark matter as of now. Therefore, we set the ratio of the energy densities between nuclear matter and dark matter, i.e., $\varepsilon_D^c / \varepsilon_N^c$, as a parameter. This method is similar to those used in ref.~\cite{Kumar25} and other studies. The radii of nuclear matter or dark matter are determined by the boundary condition of $P(R_N)=0$ or $P(R_D)=0$. If $R_N>R_D$, a dark matter core forms. A dark matter halo forms for $R_N>R_D$.}

In short, four parameters describing dark matter can vary in this model, namely $m_D$, $C_{DS}$, $C_{DV}$, and $\varepsilon_D^c / \varepsilon_N^c$. For a selected $m_D$ value, by constructing a three-dimensional (3D) parameter space of $C_{DS} - C_{DV} - \varepsilon_D^c / \varepsilon_N^c$, we can constrain these parameters simultaneously based on current observations of neutron stars, particularly the simultaneous observation of the mass and radius of PSRJ0740+6620. 

The 3D parameter space method used in this study and the well-known Bayesian analysis approach are two distinct methods for extracting observational constraints on the properties of nuclear matter or dark matter. Both are rigorous, but each has its own advantages and disadvantages. The 3D parameter space method systematically explores the parameter space without any prior assumptions, thus avoiding the exclusion of potentially valid parameter sets. It directly links the parameters to observational data, offering clearer insights into the effects of dark matter on the properties of DANSs. The resulting data are both informative and straightforward. However, this method faces challenges in quantifying uncertainties and struggles with models that contain numerous parameters. In this study, there are only four free parameters in the dark matter model, making the 3D parameter space method adequate for the analysis.

\section{Results and Discussions}\label{sec3}

The mass of PSR J0740+6620 is constrained to be $2.08 \pm 0.08$ M$_\odot$~\cite{Cromartie19, Fonseca21}, making it the most massive neutron star observed so far. In 2021, NICER obtained simultaneously the mass and radius of PSR J0740+6620 as $R = 13.7^{+2.6}_{-1.5}$ km~\cite{Miller21} and $R = 12.39^{+1.30}_{-0.98}$ km~\cite{Riley21} from two independent analyses of their observational data using somewhat different techniques. Information about both the mass and the radius of a massive neutron star provides tighter constraints on both the EOS of the nuclear matter and the dark matter.

In the present study, we are interested in how the simultaneous measurement of mass and radius of PSR J0740+6620 can help constrain the parameters $C_{DS}$, $C_{DV}$, $\varepsilon_D^c/\varepsilon_N^c$, and the dark matter fraction in a DANS. To achieve this goal, we calculate the constraints based on the upper and lower limits of NICER observations and find that only the lower limit of $R_{2.08} = 12.2$ km provides additional constraints on the properties of dark matter. A similar situation arises in analyzing other NICER and GW170817 observations. This is because the nuclear matter EOS used in the present work already describes well the current astrophysical observations within their uncertain ranges without considering dark matter{, namely, $M_{\rm max}=2.13$~M$_\odot$ and $R_{2.08}=12.25$ km. To satisfy the constraints of $M_{\rm max}=2.08$ M$_\odot$ and $R_{2.08}=12.2$ km in DANS, we only need to include a small amount of dark matter. This would cause only a slight change in the M-R curves, ensuring that the selected constraints are still satisfied. Thus, both $M_{\rm max}=2.08$ M$_\odot$ and $R_{2.08}=12.2$ km can be used to constrain the properties of dark matter.}

{However, for another alternative constraint from NICER, such as $R_{2.08}\leq16.3$ km from ref.~\cite{Miller21}, we find that this condition is always satisfied, regardless of how the dark matter parameters are varied. This is because the inclusion of dark matter tends to reduce the nuclear matter radius as the dark matter core exerts an inward force, pulling the nuclear matter inward. A similar situation arises for other NICER constraints and even tidal deformability. All these constraints are satisfied regardless of changes in the dark matter parameters. Therefore, $M_{\rm max}=2.08$ M$_\odot$ and $R_{2.08}=12.2$ km are the only two constraints that can effectively constrain the properties of dark matter.}

\begin{figure*}
  \includegraphics[width=13cm]{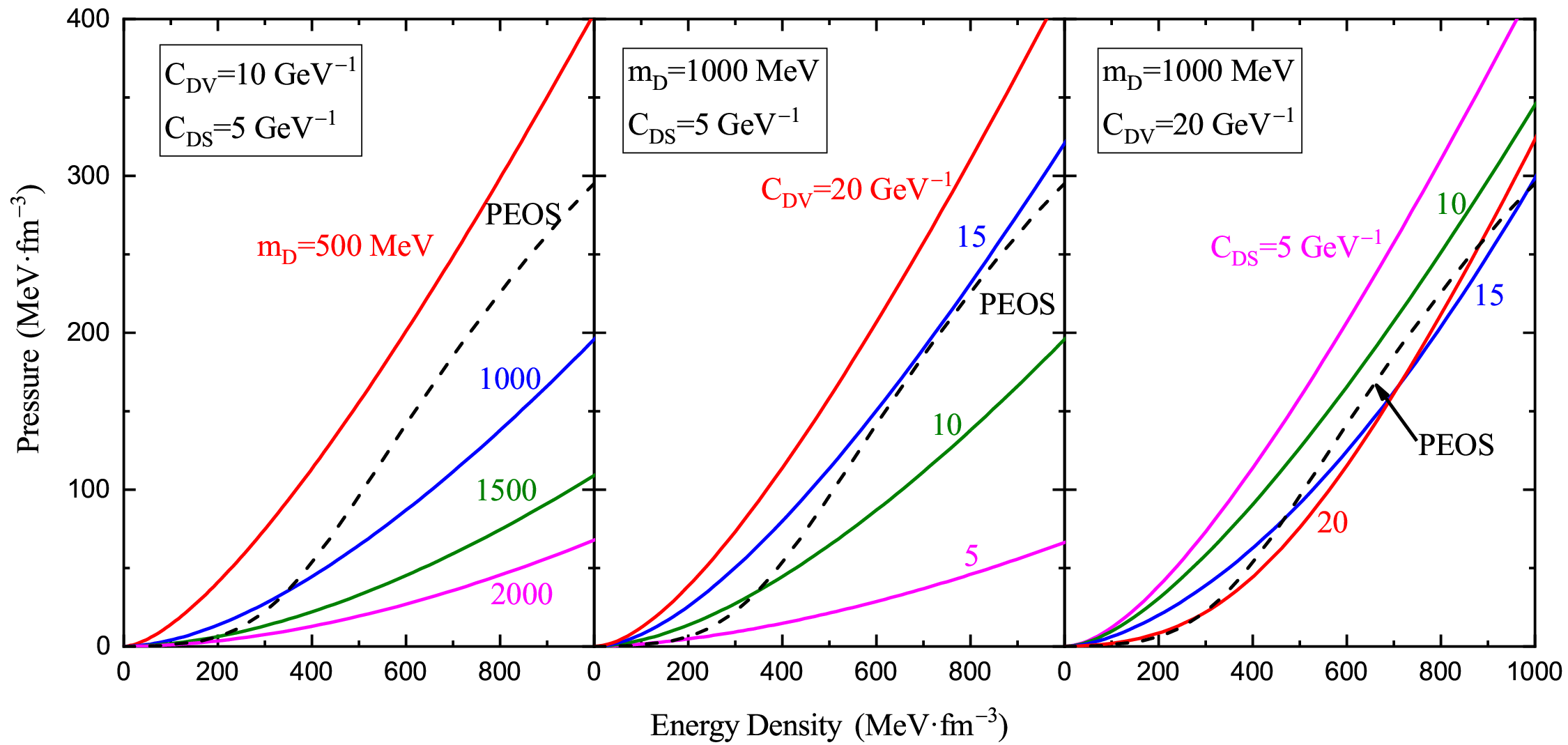}
  \caption{The effects of dark matter mass $m_D$ (left panel), $C_{DV}$ (middle panel), and $C_{DS}$ (right panel on the EOS of dark matter. The black dashed line, labeled as PEOS, is the neutron star EOS mentioned in Section~\ref{2.1} and is given for comparison.}\label{EOS}
\end{figure*}

The effects of dark matter mass $m_D$ (left panel), $C_{DV}$ (middle panel), and $C_{DS}$ (right panel) on the dark matter EOS are shown in Figure~\ref{EOS}. The black dashed line, labeled as PEOS, is the nuclear matter EOS mentioned in {Section}~\ref{2.1} and is given for comparison. Note that the dark matter EOS is independent of $\varepsilon_D^c/\varepsilon_N^c$. As clearly shown, increasing $m_D$ results in a softer dark matter EOS. This behavior can be explained by the equations of dark matter EOS, specifically the last terms in Equations (\ref{energy}) and (\ref{pressure}). An increase in $m_D$ leads to a larger energy density $\varepsilon$ but a smaller pressure $p$, resulting in a softer EOS. Compared to the PEOS of nuclear matter, the dark matter EOS becomes softer at high densities, except for the EOS corresponding to $m_D = 2000$ MeV. The contributions of $C_{DV}$ and $C_{DS}$ to the EOS are opposite, as they have opposite effects on the potential, as shown in Equation (\ref{potential}). {In addition, we can see that the EOS for $C_{DS}=20$ and 15 GeV$^{-1}$ cross around 700 MeV$\cdot$fm$^{-3}$. This is because the changes in $C_{DS}$ affect both the first and third terms of \mbox{Equations (\ref{energy}) and (\ref{pressure})}. More specifically, $C_{DS}$ also influences the effective mass $m^*$ of dark matter. As the first term plays an increasingly dominant role in determining EOS stiffness at high densities and it contributes positively to the dark matter EOS, the EOS for $C_{DS}=20$ GeV$^{-1}$ becomes stiffer at high densities. }

\begin{figure*}
   \resizebox{0.95\textwidth}{!}{
  \includegraphics[bb=0 0 600 530]{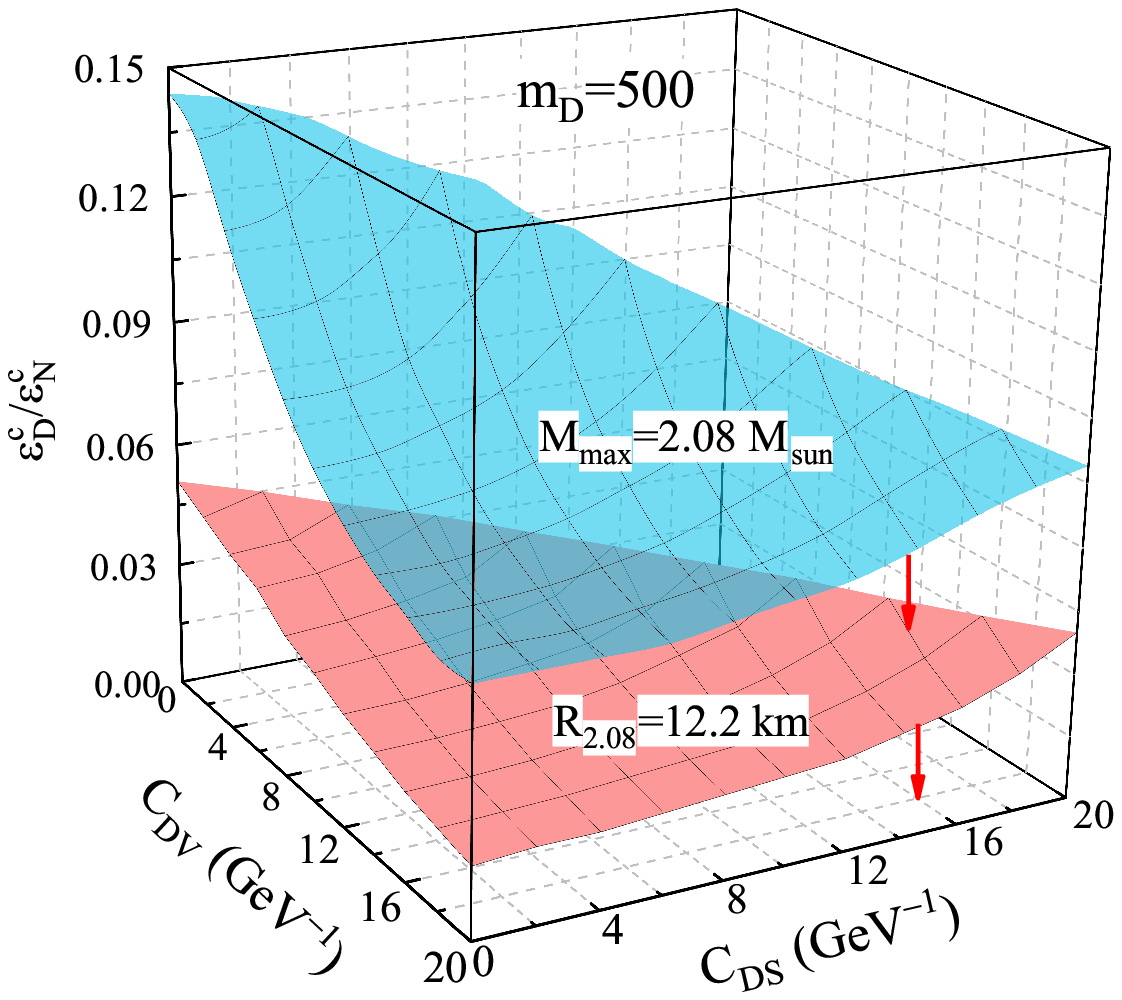}
  \includegraphics[bb=0 0 600 530]{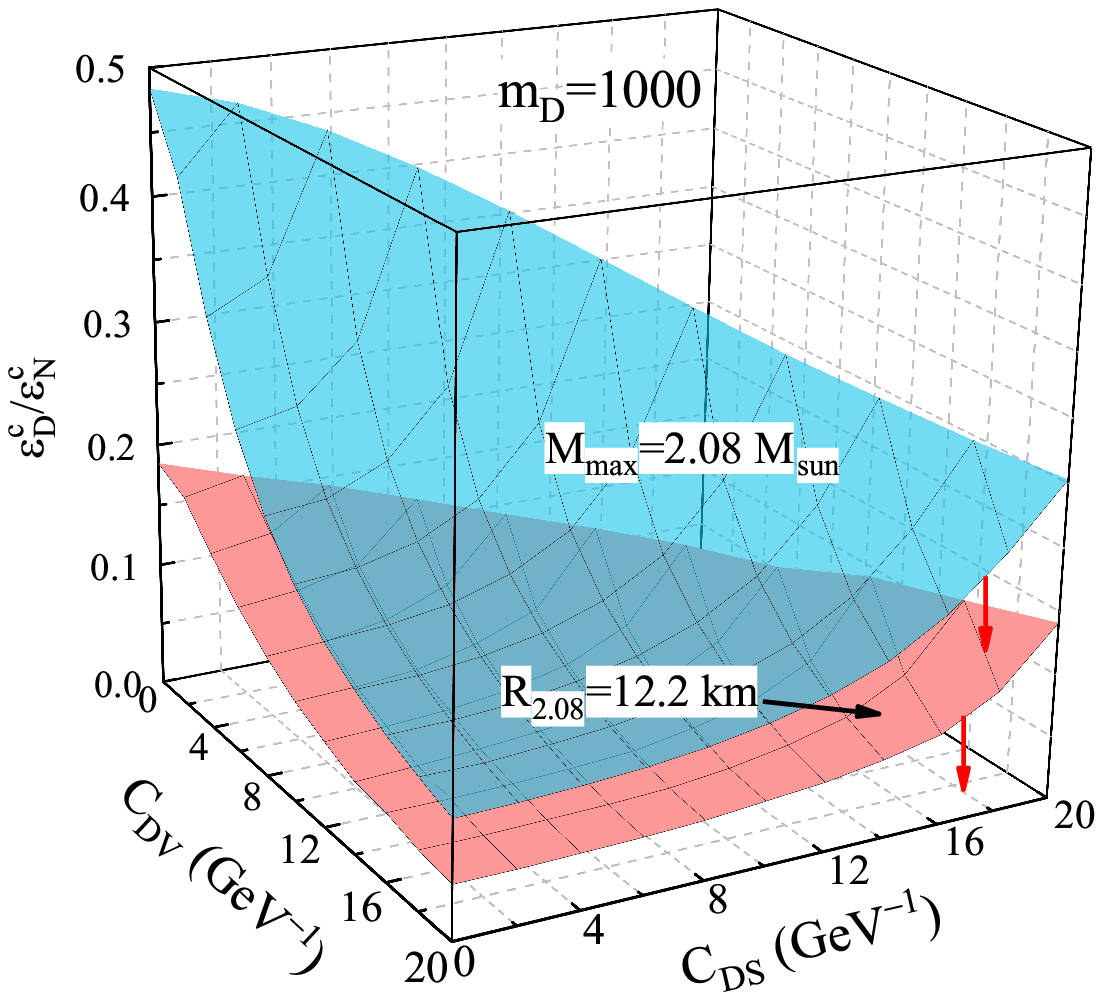}
  }\\
     \resizebox{0.95\textwidth}{!}{
  \includegraphics[bb=0 0 600 530]{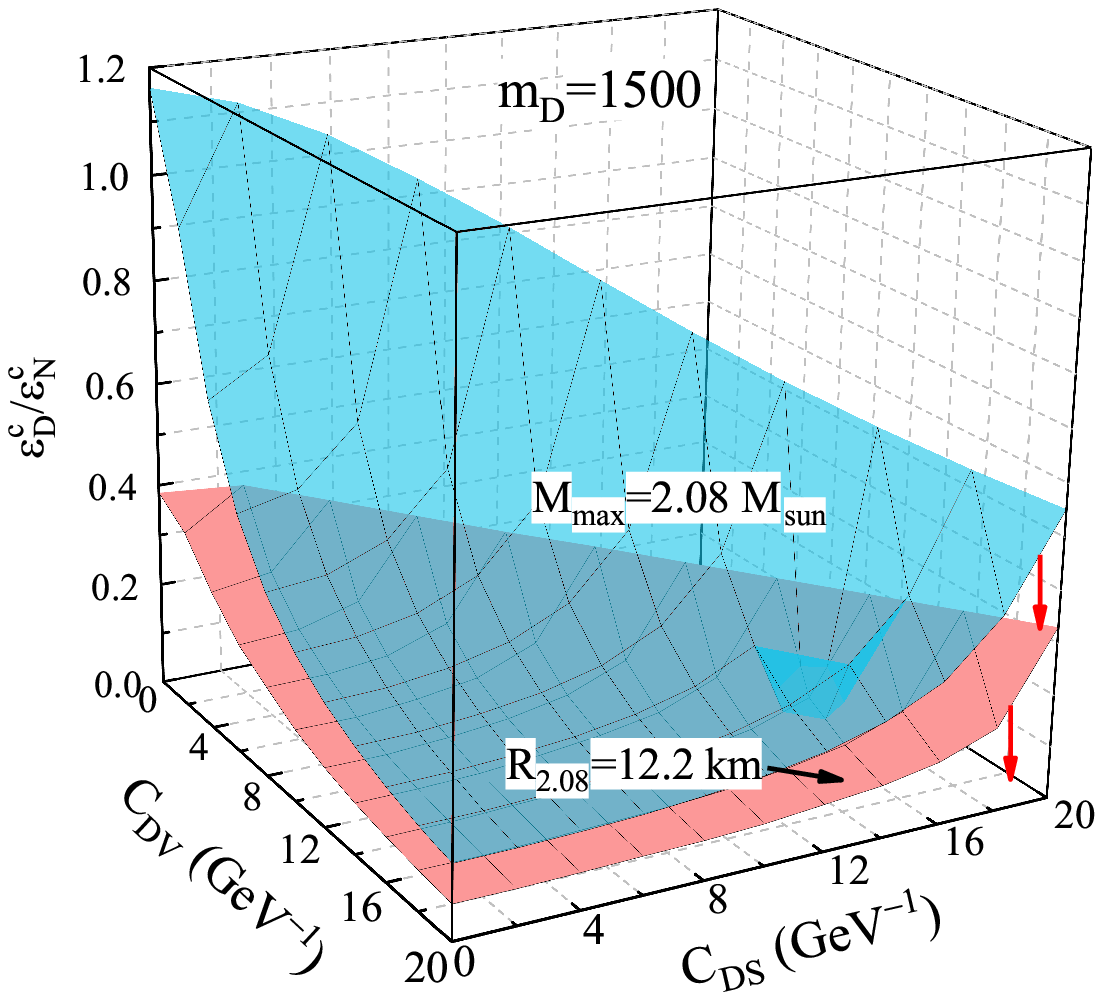}
  \includegraphics[bb=0 0 600 530]{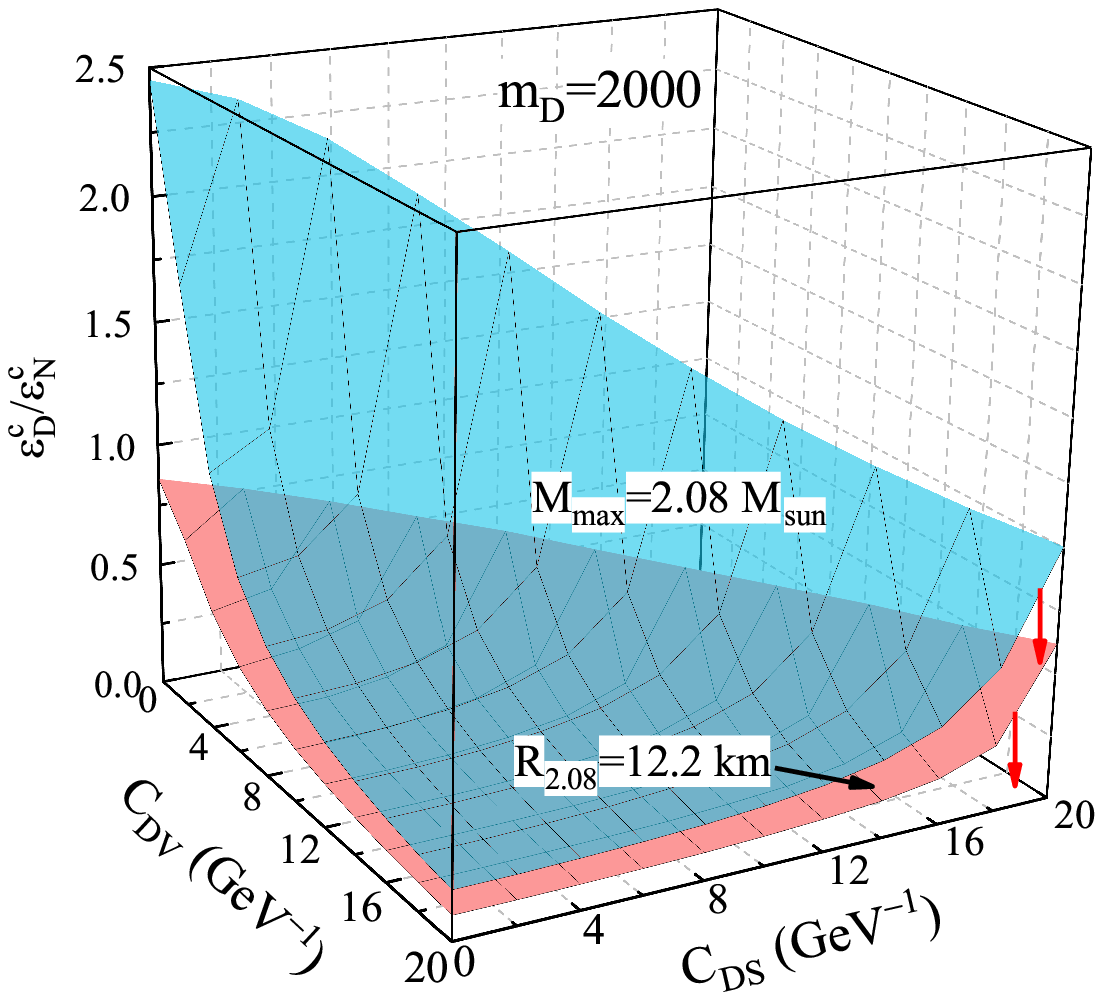}
  }
  \caption{The constant surfaces of $M_{\rm max}=2.08$ M$_\odot$ (blue surfaces) and $R_{2.08}=12.2$ km (pink surfaces) in the 3D parameter space of $C_{DS}-C_{DV}-\varepsilon_D^c/\varepsilon_N^c$ for $m_D=500$, 1000, 1500, and 2000 MeV, respectively. The red arrows indicate the directions that satisfy the corresponding observation, while the black arrow points to the surface of $R_{2.08}=12.2$ km.}\label{3D}
\end{figure*}

After understanding how each parameter affects the EOS, we can now constrain these parameters based on astrophysical observations. The constant surfaces of $M_{\rm max} = 2.08$ M$_\odot$ (blue surfaces) and $R_{2.08} = 12.2$ km (pink surfaces) in the 3D parameter space of $C_{DS} - C_{DV} - \varepsilon_D^c/\varepsilon_N^c$ for $m_D = 500$, 1000, 1500, and 2000 MeV are shown in Figure~\ref{3D}. All surfaces end at $C_{DS} = C_{DV}$. The red arrows indicate the directions that satisfy the corresponding observation, while the black arrow points to the corresponding surface. Note that the ranges of the $\varepsilon_D^c/\varepsilon_N^c$ axis differ across the four panels {and the two surfaces do not intersect in each plot}. The constant surfaces mean that any point selected from the surface corresponds to a parameter set that satisfies the corresponding observation. For example, any point on the constant surface of $M_{\rm max} = 2.08$ M$_\odot$ corresponds to a parameter set ($C_{DS}$, $C_{DV}$, $\varepsilon_D^c/\varepsilon_N^c$) that leads to an EOS of DANSs with a maximum mass of $2.08$ M$_\odot$. Therefore, the constant surfaces in the 3D parameter space help constrain the parameters.
\vspace{-6pt}

From Figure~\ref{3D}, we can see that the observed mass of PSR J0740+6620 can significantly constrain the parameter space, and the simultaneous measurement of its mass and radius provides even tighter constraints. However, only the parameter $\varepsilon_D^c/\varepsilon_N^c$ can be directly constrained, while $C_{DS}$ and $C_{DV}$ remain unconstrained. This is because there is no direct constraint on $\varepsilon_D^c/\varepsilon_N^c$ at present. For instance, for $m_D = 2000$ MeV, if $\varepsilon_D^c/\varepsilon_N^c$ is constrained to be less than 1, we can plot a horizontal constant surface of $\varepsilon_D^c/\varepsilon_N^c = 1$ to intersect other parameter surfaces in the space. The crossline between $\varepsilon_D^c/\varepsilon_N^c = 1$ and \mbox{$M_{\rm max} = 2.08$ M$_\odot$} provides a projection onto the $C_{DS} \sim C_{DV}$ plane, which helps constrain $C_{DS}$ or $C_{DV}$. Despite $C_{DS}$ and $C_{DV}$ being unconstrained by the current observations, we obtain a reliable constraint on $\varepsilon_D^c/\varepsilon_N^c$. The upper limit for $\varepsilon_D^c/\varepsilon_N^c$ increases from 0.15 to 2.45 as $m_D$ increases from 500 to 2000 MeV for $M_{\rm max} = 2.08$ M$_\odot$. This is because the dark matter EOS becomes softer for larger $m_D$, requiring more dark matter and thus a larger $\varepsilon_D^c$ to support a DANS with $M_{\rm max} = 2.08$ M$_\odot$. Additionally, all constant surfaces of $M_{\rm max} = 2.08$ M$_\odot$ provide the strictest constraints on $\varepsilon_D^c/\varepsilon_N^c$ at $C_{DS} = 0$ and $C_{DV} = 20$ GeV$^{-1}$. This is consistent with Figure~\ref{EOS}, where the combination of $C_{DS} = 0$ and $C_{DV} = 20$ GeV$^{-1}$ results in the {hardest} dark matter EOS, requiring the least dark matter to support a DANS with \mbox{$M_{\rm max} = 2.08$ M$_\odot$}.

A similar trend is observed when the mass and radius of PSR J0740+6620 are measured simultaneously ($R_{2.08} = 12.2$ km), as shown in Figure~\ref{3D}. However, the constant surface of $R_{2.08} = 12.2$ km reduces the upper limit of $\varepsilon_D^c/\varepsilon_N^c$ by more than 50\% for any $m_D$. The upper limit decreases to only 0.05 for $m_D = 500$ MeV and 0.86 for $m_D = 2000$ MeV. Using the constraints from~\cite{Das22b} for $C_{DS} = 3.90$ GeV$^{-1}$ and $C_{DV} = 11.88$ GeV$^{-1}$, we find $\varepsilon_D^c/\varepsilon_N^c = 0.07$, 0.17, 0.26, and 0.37 for $m_D = 500$, 1000$, 1500$, and 2000 MeV, respectively. This suggests that lighter dark matter mass corresponds to a smaller $\varepsilon_D^c/\varepsilon_N^c$ in DANSs. However, lighter dark matter mass leads to a larger dark matter fraction, as discussed below.

\begin{figure*}
  
  \includegraphics[width=13cm]{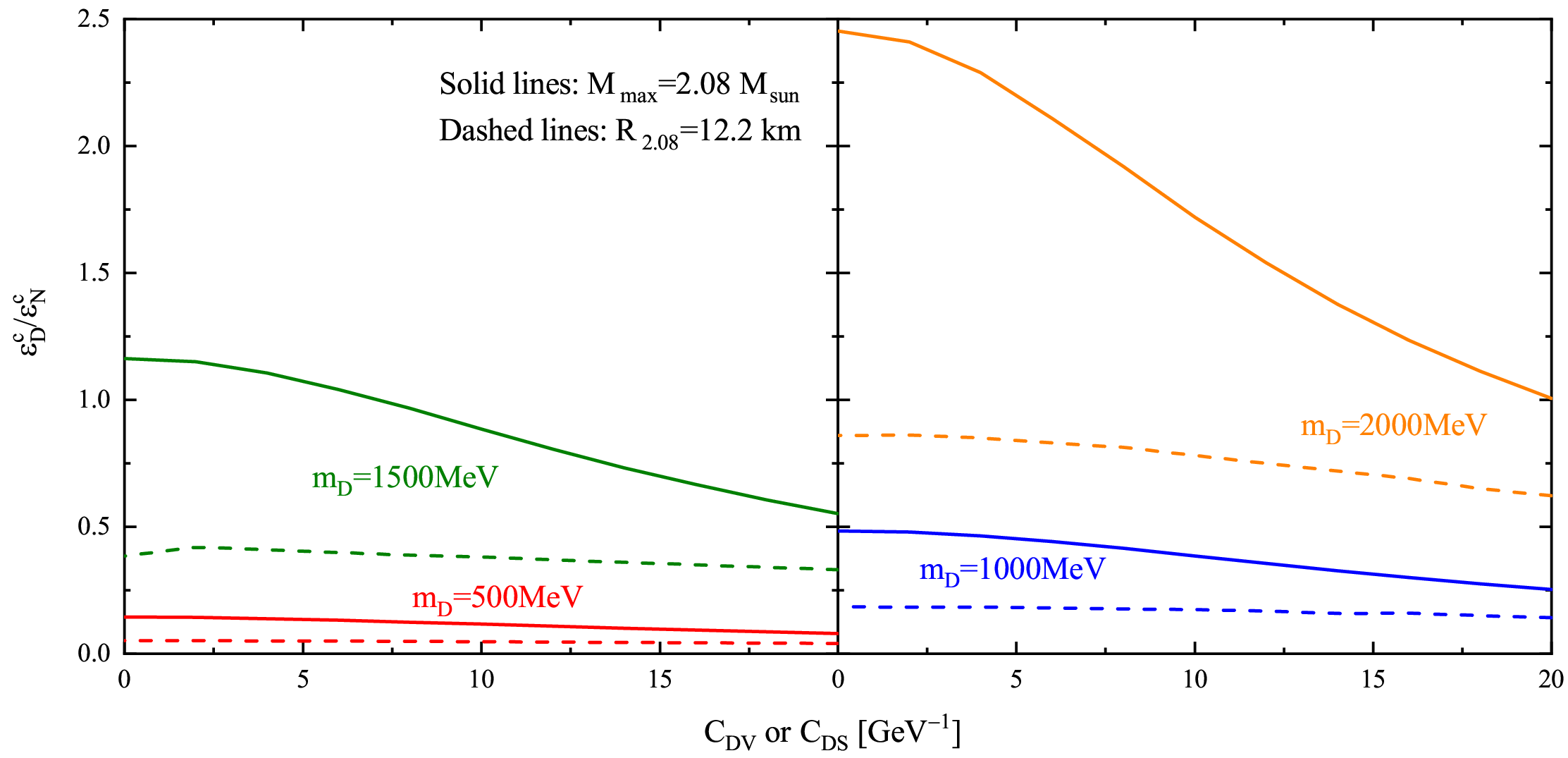}
  \caption{The projection of $C_{DS}=C_{DV}$ to the $C_{DS}\sim\varepsilon_D^c/\varepsilon_N^c$ or $C_{DV}\sim\varepsilon_D^c/\varepsilon_N^c$ plane with \mbox{$m_D$ = 500} (red lines), 1000 (blue lines), 1500 (green lines), and 2000 (orange lines) MeV for the surfaces of $M_{\rm max}=2.08$ M$_\odot$ (solid lines) and $R_{2.08}=12.2$ km (dashed lines).}\label{Rho}  
\end{figure*}

For both $M_{\rm max} = 2.08$ M$_\odot$ and $R_{2.08} = 12.2$ km, $C_{DS} = C_{DV}$ represents the upper boundary for the constant surfaces. Therefore, by projecting the line of $C_{DS} = C_{DV}$ along each surface to the $C_{DS} \sim \varepsilon_D^c/\varepsilon_N^c$ plane, we can better visualize the constraints on $\varepsilon_D^c/\varepsilon_N^c$. Note that the projections to the $C_{DS} \sim \varepsilon_D^c/\varepsilon_N^c$ or $C_{DV} \sim \varepsilon_D^c/\varepsilon_N^c$ plane are equivalent, as $C_{DS} = C_{DV}$ is satisfied for any surface. To make the constraints on $\varepsilon_D^c/\varepsilon_N^c$ clearer, the projection of $C_{DS} = C_{DV}$ onto the $C_{DS} \sim \varepsilon_D^c/\varepsilon_N^c$ plane is shown for $m_D = 500$ (red lines), $1000$ (blue lines), $1500$ (green lines), and $2000$ (orange lines) MeV. These projections are presented for the surfaces of $M_{\rm max} = 2.08$ M$_\odot$ (solid lines) and $R_{2.08} = 12.2$ km (dashed lines) in Figure~\ref{Rho}.

For the solid lines corresponding to $M_{\rm max} = 2.08$ M$_\odot$, the dependence of $\varepsilon_D^c/\varepsilon_N^c$ on $C_{DS}$ or $C_{DV}$ exists for $m_D = 2000$ and $1500$ MeV but decreases as $m_D$ decreases. This dependence disappears as the lines become much flatter for $m_D = 1000$ and $500$ MeV. This behavior can be explained from Figure~\ref{3D}. Since the maximum mass of the PEOS is 2.13~M$_\odot$ and the existence of dark matter reduces the maximum mass, a small amount of dark matter must be introduced to obtain a star with $M_{\rm max} = 2.08$ M$_\odot$. Based on the stiffness of the dark matter EOS, dark matter with smaller $m_D$ requires a smaller central density to contribute similarly to dark matter with larger $m_D$. Therefore, tighter constraints on $\varepsilon_D^c/\varepsilon_N^c$ are obtained for smaller $m_D$.

However, when considering the constraint of $R_{2.08} = 12.2$ km, we observe that the density of dark matter must always be smaller than that of nuclear matter for different values of $m_D$. Furthermore, all the dashed lines appear horizontal, indicating that $\varepsilon_D^c/\varepsilon_N^c$ is independent of $C_{DS}$ or $C_{DV}$ for any value of $m_D$. Meanwhile, the constrained value of $\varepsilon_D^c/\varepsilon_N^c$ decreases with decreasing $m_D$. This suggests that we cannot obtain additional constraints on $C_{DS}$ and $C_{DV}$ through the simultaneous measurement of the mass and radius of DANSs. Therefore, the simultaneous measurement of the mass and radius of DANSs can help constrain $\varepsilon_D^c/\varepsilon_N^c$ but not $C_{DS}$ and $C_{DV}$, even if tighter constraints are obtained in the~future.

\begin{figure*}
  
  \includegraphics[width=12cm]{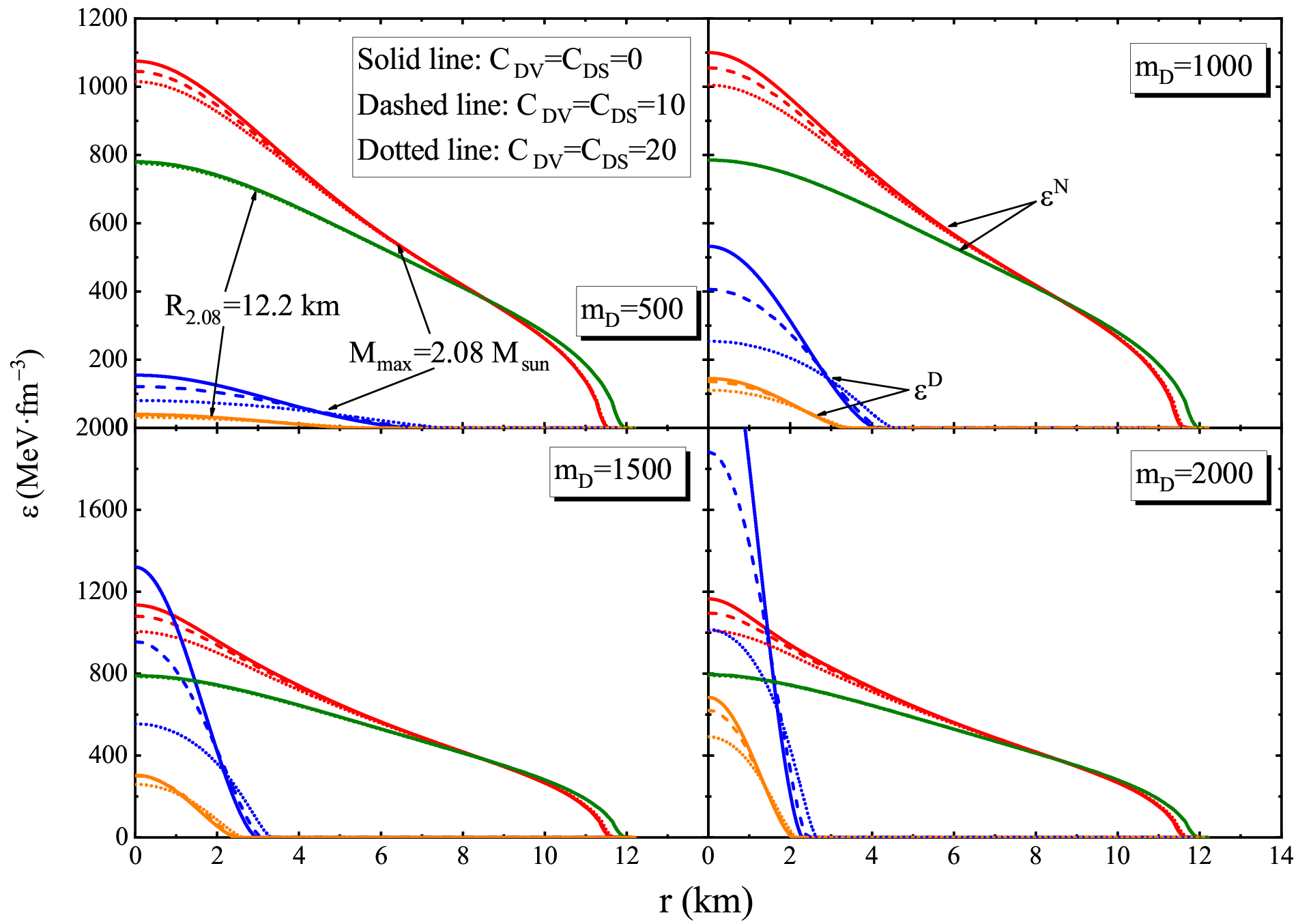}
  \caption{The profile of energy density as functions of distance $r$ for \mbox{$M_{\rm max}=2.08$ M$_\odot$} and \mbox{$R_{2.08}=12.2$ km} and $C_{DS}=C_{DV}$ = 0, 10, 20 GeV$^{-1}$, respectively. The red and green lines correspond to the profiles of nuclear matter, while the blue and orange lines correspond to the profile of dark matter.}\label{Proec}
\end{figure*}

In the previous discussions, we showed how observations can constrain the properties of dark matter. However, the amount of dark matter in a DANS cannot be determined solely by knowing the central density. Thus, a profile of the density as a function of the distance from the center of the star is needed. To plot this profile, we select three representative parameter sets from Figure~\ref{Rho}, specifically $C_{DS} = C_{DV} = 0$, 10, and 20 GeV$^{-1}$, to illustrate the possible effects of varying $C_{DS}$ and $C_{DV}$ on the density profile. Since the projections in Figure~\ref{Rho} are monotonic or relatively flat, these three combinations of $C_{DS}$ and $C_{DV}$ capture the main features of the dark matter properties. The energy density profile as functions of distance $r$ for $M_{\rm max} = 2.08$ M$_\odot$ and $R_{2.08} = 12.2$ km, with $C_{DS} = C_{DV} = 0$, 10, and 20~GeV$^{-1}$, are shown in Figure~\ref{Proec}. The red and green lines represent the profiles of nuclear matter, while the blue and orange lines represent the dark matter profiles. Note that the profiles are displayed for stars with $M = 2.08$ M$_\odot$.

As expected, only the dark matter core exists for the selected values of $m_D$. We observe that the effects of $C_{DS}$ and $C_{DV}$ on the profile are noticeable only for $M_{\rm max} = 2.08$ M$_\odot$ but disappear for $R_{2.08} = 12.2$ km. For $M_{\rm max} = 2.08$ M$_\odot$, $C_{DS}$ and $C_{DV}$ affect only the central part of the star, while they have almost no effect on the core for $R_{2.08} = 12.2$ km. This is because the central density remains the same for different values of $C_{DS}$ and $C_{DV}$ for $R_{2.08} = 12.2$ km, as shown in Figure~\ref{Rho}; therefore, they do not impact the density~profiles.

In addition, the size of the dark matter core decreases as $m_D$ increases. For \mbox{$M_{\rm max} = 2.08$ M$_\odot$}, the size of the dark matter core decreases from 8.00 km to 2.41 km as $m_D$ increases from 500 MeV to 2000 MeV when $C_{DS} = C_{DV} = 0$. The change in $C_{DS}$ and $C_{DV}$ from 0 to 20 GeV$^{-1}$ leads to a correction of less than 1 km but significantly alters the values of $\varepsilon_D^c$ and $\varepsilon_N^c$. Although $\varepsilon_D^c$ can exceed 2000 MeV$\cdot$fm$^{-3}$, the small size of the dark matter core may suppress the total amount of dark matter in the DANS. In other words, the competition between the core's size and the central density determines the amount of dark matter in the DANS.

To quantify the amount of dark matter in the DANS, we define the fraction of dark matter as
\begin{equation}
 f_D=M_D/M_t,
\end{equation}
where $M_D$ is the total mass of dark matter and $M_t$ is the total mass of the DANS within distance $r$. Currently, $f_D$ is not well constrained. Bayesian analyses, such as ref.~\cite{Liu25}, found that $f_D = 1.3$$\sim$7.9\%, while other studies~\cite{Arvikar25, Das22b} propose ranges of $f_D = 0.8$$\sim$5.9\% and $f_D = 6$$\sim$11\%. Additionally, ref.~\cite{Thakur24b} found $f_D = 10$$\sim$25\% with the constraint of \mbox{$M_{\rm max} > 1.9$ M$_\odot$}, and refs.~\cite{Karkevandi22, Ellis18} report $f_D \approx 5\%$.

\begin{figure}
  
  \includegraphics[width=7.7cm]{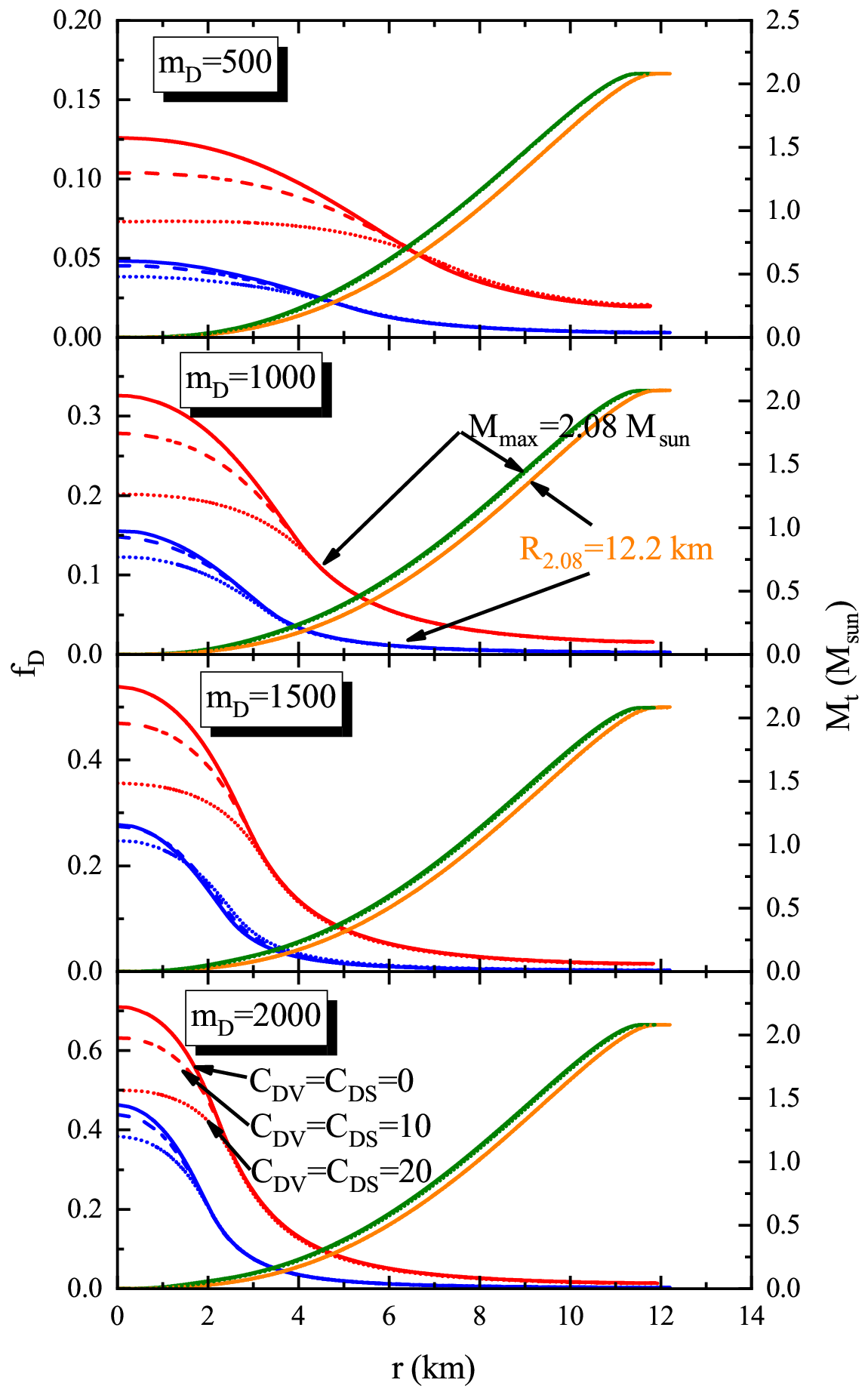}
  \caption{The profile of dark matter fraction $f_D$ (left title, red and blue lines) and total mass $M_t$ (right title, green and orange lines) as functions of distance $r$ for $M_{\rm max}=2.08$ M$_\odot$ and $R_{2.08}=12.2$ km and $C_{DS}=C_{DV}$ = 0, 10, 20 GeV$^{-1}$, respectively.}\label{Promass} 
\end{figure}

To constrain $f_D$ based on the present observations of \mbox{$M_{\rm max} = 2.08$ M$_\odot$} and\linebreak \mbox{$R_{2.08} = 12.2$ km}, the profiles of dark matter fraction $f_D$ (left panel, red and blue lines) and total star mass $M_t$ (right panel, green and orange lines) as functions of distance $r$ for $C_{DS} = C_{DV} = 0$, 10, and 20 GeV$^{-1}$ are shown in Figure~\ref{Promass}. Note that the ranges of the $f_D$ axis differ for different values of $m_D$. It is clear that $f_D$ decreases with distance $r$. For $m_D = 500$ MeV, the largest $f_D$ is 12.6\% for $M_{\rm max} = 2.08$ M$_\odot$ and 4.8\% for $R_{2.08} = 12.2$ km when $C_{DS} = C_{DV} = 0$. Again, the effects of $C_{DS} = C_{DV}$ are visible only in the central region of the star. Although $f_D$ is relatively large at the center of the star, it drops to lower values when $r \approx 6$ km, where the star's mass is still less than 0.5 M$_\odot$. Therefore, the total $f_D$ is not expected to be large for the entire star. At $r = R$, $f_D$ quickly decreases to 1.96\% for $M_{\rm max} = 2.08$ M$_\odot$ and 0.3\% for $R_{2.08} = 12.2$ km, and the effects of $C_{DS} = C_{DV}$ on $f_D$ disappear. For $m_D = 2000$ MeV, $f_D$ is constrained to be 1.4\% for $M_{\rm max} = 2.08$ M$_\odot$ and 0.3\% for $R_{2.08} = 12.2$ km.

Thus, we conclude that the simultaneous measurement of the mass and radius of PSRJ0740+6620 can constrain the dark matter fraction $f_D$ to around 0.3\%. Tighter constraints may support smaller values of $\varepsilon_D^c/\varepsilon_N^c$ and, consequently, a smaller dark matter fraction $f_D$.

\section{Conclusions}\label{sec4}

In this study, {we investigate how the observations from NICER can help constrain the properties of dark matter in DANSs. We assume the dark matter to be fermionic dark matter, where dark matter interacts exclusively via gravity. Both vector and scalar fields are considered in the fermionic dark matter. For the nuclear matter in DANSs, we adopt a parameterized EOS from ref.~\cite{Zhang18}. All parameters have been constrained based on current terrestrial nuclear experiments and neutron star observations. We only utilize this EOS to describe nuclear matter to reduce the complexity arising from the uncertainties in the nuclear matter EOS.} We compare the constraints on dark matter properties within the DANS by considering both the observed neutron star maximum mass ($M_{\rm max} = 2.08$ M$_\odot$) and the simultaneous measurement of mass and radius ($R_{2.08} = 12.2$ km) for PSRJ0740+6620 in the 3D parameter space of $C_{DS} - C_{DV} - \varepsilon_D^c / \varepsilon_N^c$. The nuclear matter EOS selected in this study satisfies all relevant observational constraints, including the maximum neutron star mass as well as the masses and radii of several neutron stars observed by NICER and GW170817 by LIGO/VIRGO.

Our results indicate that the simultaneous measurement of the mass and radius can reduce uncertainty in $\varepsilon_D^c / \varepsilon_N^c$ by more than 50\%. However, the parameters $C_{DS}$ and $C_{DV}$ remain unconstrained, even with the potential for tighter constraints from future measurements. To better constrain these parameters, new kinds of observation will be essential. Although the central energy density of dark matter can be much larger than that of nuclear matter, the dark matter fraction $f_D$ in the star should remain below 2\% with the $M_{\rm max} = 2.08$ M$_\odot$ constraint and could be as low as 0.3\% with the $R_{2.08} = 12.2$ km constraint. This suggests that only a small amount of dark matter is present in the DANS.

The nuclear matter EOS we selected is consistent with all existing astrophysical constraints, making it relatively easy for DANSs to satisfy these conditions, as the needed changes from dark matter EOS to the nuclear matter EOS are minimal. Essentially, there is a degeneracy between the EOS of nuclear matter and that of dark matter because the TOV equations are composition-blind. 
Namely, different combinations of nuclear matter and dark matter can lead to the same pressure and energy density necessary to describe the observations. Naturally, other choices of nuclear matter EOSs could lead to more stringent constraints on dark matter properties. The inclusion of new particles, such as hyperons and mesons, or the consideration of a phase transition from nuclear matter to quark matter could also modify the nuclear matter EOS. Additionally, we focused on the lower limit $R_{2.08} = 12.2$ km constraint derived from the simultaneous measurement of mass and radius. We did not discuss the corresponding upper limit of $R_{2.08} = 16.3$ km or the analysis from Ref.~\cite{Riley21}, as these are satisfied by almost all EOS parameter combinations within our model framework. Therefore, tighter constraints on the masses and especially radii of massive neutron stars are crucial to deepen our understanding of dark matter properties in DANSs.

\noindent{\bf Acknowledgments}\\
We would like to thank Wei-Zhou Jiang for helpful discussions. BAL is supported in part by the U.S. Department of Energy, Office of Science, under Award Number DE-SC0013702, the CUSTIPEN (China-U.S. Theory Institute for Physics with Exotic Nuclei) under the US Department of Energy Grant No. DE-SC0009971. NBZ is supported in part by the National Natural Science Foundation of China under Grant No. 12375120 and the Zhishan Young Scholar of Southeast University under Grant No. 2242024RCB0013. JYZ, WNS, and HZ are supported by the National College Students' Innovation and Entrepreneurship Training Program.


\end{document}